\pdfoutput=1

\documentclass[sigconf, nonacm]{acmart}
\usepackage{booktabs}
\usepackage{array}
\usepackage{tabularx}

\usepackage{graphicx}
\usepackage{multirow}

\usepackage{algorithm,algorithmic,amsmath}

\usepackage{url}

\makeatletter
\def\algbackskip{\hskip-\ALG@thistlm}
\makeatother

\input{Definitions.tex}
\AtBeginDocument{%
  \providecommand\BibTeX{{%
    \normalfont B\kern-0.5em{\scshape i\kern-0.25em b}\kern-0.8em\TeX}}}
    
\begin{document}


\title{Embracing Structure in Data for Billion-Scale \\
Semantic Product Search}






\author{Vihan Lakshman}
\affiliation{Amazon}
\email{vihan@amazon.com}

\author{Choon Hui Teo}
\affiliation{Amazon}
\email{choonhui@amazon.com}

\author{Xiaowen Chu}
\affiliation{Amazon}
\email{xiaowec@amazon.com}

\author{Priyanka Nigam}
\affiliation{Amazon}
\email{nigamp@amazon.com}

\author{Abhinandan Patni}
\affiliation{Amazon}
\email{abhpat@amazon.com}

\author{Pooja Maknikar}
\affiliation{Amazon}
\email{maknp@amazon.com}

\author{S.V.N. Vishwanathan}
\affiliation{Amazon}
\email{vishy@amazon.com}

\renewcommand{\shortauthors}{Lakshman, et al.}

\begin{abstract}

  We present principled approaches to train and
  deploy \emph{dyadic} neural embedding models at the billion scale, 
  focusing our investigation on the application of semantic
  product search. When training a dyadic model, one seeks to embed two
  different types of entities (e.g., queries and documents or users and
  movies) in a common vector space such that pairs with high relevance
  are positioned nearby. During inference, given an
  embedding of one type (e.g., a query or a user), one
  seeks to retrieve the entities of the other type (e.g., documents or
  movies, respectively) that are highly relevant. In this work, 
  we show that exploiting the natural structure of real-world
  datasets helps address both challenges efficiently. Specifically, we
  model dyadic data as a bipartite graph with edges between pairs with
  positive associations. We then propose to partition this network into
  semantically coherent clusters and thus reduce our search space by focusing on
  a small subset of these partitions for a given input. During training, 
  this technique enables us to efficiently mine hard negative examples
  while, at inference, we can quickly find the nearest neighbors for a given
  embedding. We provide offline experimental
  results that demonstrate the efficacy of our techniques for both
  training and inference on a billion-scale Amazon.com product search dataset.

\end{abstract}



\maketitle
\pagestyle{plain}

\section{Introduction}
\label{sec:Introduction}


Many real-world problems can be modeled with the following general
paradigm: there are two different types of entities, say $\Qcal$ and
$\Dcal$, and one observes positive interactions between pairs, say
$\rbr{q, d^{+}}$ where $q \in \Qcal$ and $d^{+} \in \Dcal$. In some
cases, negative interactions or contextual information about the
interactions are also observed. Given this, so-called, \emph{dyadic}
data, the goal is to generalize, that is, to either a) predict new
positive interaction pairs for existing entities or b) predict the
interactions for unseen entities. By carefully selecting the sets
$\Qcal$ and $\Dcal$ and determining which interactions are considered
positive\footnote{In some cases, positive and/or negative interactions
  may be only observed implicitly.}, a variety of problems can be cast
in this framework. For instance, let $\Qcal$ be the set of queries that
users type into a search engine, and $\Dcal$ be the set of all documents
on the Internet.  Furthermore, define an interaction as positive if the
user types a query $q$ and clicks on a document $d$. The generalization
problem is to find matching documents for a query $q$. In product
search, $\Qcal$ is the set of queries and $\Dcal$ is the set of products; 
if a query $q$ was used to purchase a product $d$, then that
interaction is defined as positive. Similarly $\Qcal$ can be a set of
users and $\Dcal$ a set of movies, with $(q, d)$ being positive if
the user watched the movie and rated it highly. The generalization
problem in this case is to recommend relevant movies to a user
$q$. 


One popular approach to dealing with dyadic data is to use dyadic neural
embedding models; one embeds entities $q \in \Qcal$ and $d \in \Dcal$
into a common vector space (say the $l$-dimensional Euclidean
space $\RR^{l}$) such that $\inner{\qb}{\db}$ is high for pairs with positive
interactions. Here, we used the notation $\qb$ (respectively $\db$) to
denote the $l$-dimensional embedding of $q$ (respectively $d$), and
$\inner{\cdot}{\cdot}$ to denote the usual Euclidean dot product. As the
name implies, neural models use a deep neural network to represent the
embedding function $f\rbr{\cdot}$ which maps $q \to f_q\rbr{q} := \qb$ and
$d \to f_d\rbr{d} := \db$.

As web-scale data is becoming ubiquitous, there is a growing desire to
train and deploy such models at massive scale involving hundreds of millions or even
billions of entities and interactions. Many applications also require real-time inference  
with latencies on the order of tens or hundreds of milliseconds. We find that 
the same underlying technique happens to address both these problems 
by leveraging the structure inherent in real-world data. In this
paper, we study the problems of training and deploying neural
embedding models operating on dyadic data with over a 
billion of entities, focusing our investigation on product search
where the underlying task is to retrieve relevant items from a large
catalog for a given search query.

To understand the challenge in training dyadic neural embedding models,
note that the models require supervision with both positive and negative
examples. As noted above, positive pairs, such as clicked
query-document pairs, are determined by the underlying task and
typically only form a miniscule fraction of all possible pairs. Negative
examples, on the other hand, constitute a much larger set since most
pairs in the universe of possibilities are dissimilar. On large
datasets, selecting random pairs as negatives proves to be too
\emph{easy} as the probability of randomly chosen items exhibiting high
dot-product values in the embedding space becomes minuscule. Instead, we
desire tuples of related, but ultimately dissimilar entities. Such
\emph{hard} negative examples induce a larger loss and thereby produce
more effective parameter updates. These hard negative examples 
can improve model generalization and accelerate convergence, both in terms
of wall clock time and in sample complexity. However, efficiently identifying such
informative negative examples emerges as a challenge for larger datasets since it becomes
computationally infeasible to examine all possible pairs.

Moreover, given a query embedding $\qb$, finding document
embeddings $\db$ with large dot product value $\inner{\qb}{\db}$ remains
the essence of deploying dyadic embedding models where the problem is to
identify the $k$-nearest neighbors of $\qb$ from the set of embedded
documents.

As can be seen, both training and inference boil down to the problem of
finding nearby points in the embedding space. If one had access to an
efficient oracle to solve this problem, then both training and inference
of dyadic neural models could be scaled up. In this paper, we present a
technique for approximating such an oracle by leveraging the fact that
real-world dyadic data is highly structured, often exhibiting a
fine-grained separability. For instance, in product search, the set of
queries that lead to the purchase of diapers do not overlap
with queries used to buy shoes.

In a nutshell, we model dyadic data as a bipartite graph with edges
between positively associated pairs. We partition the nodes of this graph
into balanced clusters by approximately
minimizing edge-cuts (the number of edges that cross cluster
boundaries). Given this partitioned graph, we can speed up training and
inference as follows: 
\begin{description}
\item[Training: ] Given a positive example $\rbr{q, d^+}$, we sample items in $\Dcal$
  from graph clusters adjacent to the one containing $q$ to find
  hard negative examples of the form $(q, d^-)$ to add to a mini-batch
  during training.
\item[Inference: ] Given an input embedding $\qb$, we seek to find $k$
  points in the set $\Dcal$ whose embeddings are closest to $\qb$ in the
  embedding space. For large datasets, an exact search examining all
  pairs becomes infeasible under the strict latency constraints of 
  industrial production systems. Moreover, approximate nearest neighbors
  algorithms, while dramatically reducing the latency at the expense of 
  diminished recall, typically require a time overhead in building 
  an index to search over. This index build time presents a deployment challenge in
  real-world search systems where indexes are often rebuilt with updated data
  on a daily basis. Instead, we train a classifier that predicts
  the clusters most likely to contain the nearest neighbors to $\qb$ and
  then perform a search only within those clusters using \emph{any} popular
  nearest neighbor search algorithm as a subroutine within the partitioning. For some approximate
 algorithms, we can reduce the index build time considerably since the indexes for the partitions can be constructed in parallel. Furthermore, 
 for other classes of approximate algorithms, we can also reduce the search latency.
 Moreover, we can use our classifier to assign new documents to clusters and thereby also avoid re-running our graph partitioning 
 step from scratch. 
 These improvements enable us to deploy approximate well-known nearest neighbor algorithms in a production setting. 
\end{description}

Our contributions can be summarized as follows:
\begin{itemize}
\item We propose a data-dependent algorithm that exploits the natural
  structure inherent in real-world datasets to model dyadic data as a
  bipartite graph, which in turn is partitioned to approximately
  minimize edge-cut. When applied to the product search task, the 
  partitioned graph is used to find hard negative
  examples on the fly during training. This considerably speeds up training and
  improves the generalization capability of the final model.
\item We propose to learn a classifier which learns to predict the
  clusters which are likely to contain the nearest neighbors of an
  embedded query. This allows us to limit the search for nearest
  neighbors to a small subset of the documents, speeding up
  inference and index build times by orders of magnitude. 
 \item Our work benchmarks nearest 
 neighbor algorithms at the billion scale under constraints representative of a real production
 product search system. In particular, we search over queries one-by-one as opposed to 
 batching which adds unnecessary delay in the response to queries in a real-time system.
 Secondly, we retrieve 100 items for each query as opposed to 1 since product
 search systems often involve retrieving a larger set of results to produce a more satisfying shopping experience. Finally, we report
 and analyze the index build time of approximate nearest neighbor algorithms as 
 another key metric to influence tradeoff decisions. 
\item We demonstrate the scaling behavior of our algorithms on a billion-scale
product search dataset, providing offline experiments
demonstrating the feasibility of embedding-based retrieval for semantic
product search at this scale. In particular we show that our methods lead to a faster time to convergence
during training and improved generalizability. For inference, we provide
 a general algorithmic primitive  to scale both exact and approximate $k$-nearest neighbor (KNN)
 algorithms such as HNSW, NGT, and inverse file index (IVF) methods
 along the dimensions of latency and index build time. Additionally, we perform our KNN search on a CPU
 machine and avoid the need for GPUs or other types of specialized hardware for inference. 
\end{itemize}

Our contributions also add to the growing body of work showing that
data-dependent algorithms, which take advantage of the specific
structure in a given dataset, can dramatically outperform data-independent
algorithms that guard against the worst case. 

The rest of the paper is structured as follows: in Section~\ref{sec:Background} 
we provide a brief background about factorized dyadic embedding models,
which are used to illustrate our ideas on scaling up training and
inference. Our algorithms are described in
Section~\ref{sec:SpeedingupTraining}. We place our contributions in the context
of related work in Section~\ref{sec:RelatedWork}. Experimental results
can be found in Section~\ref{sec:Experiments}, and we conclude with a
brief discussion of future work in
Section~\ref{sec:ConclusionFutureWork}.  

\section{Background}
\label{sec:Background}

As in \cite{hofmann1999learning} we define \emph{dyadic data} as a
domain with two finite sets of entities
$\Qcal = \{q_{1}, \ldots, q_{n}\}$ and
$\Dcal = \{d_{1}, \ldots, d_{m}\}$, where the set of positive
observations $p \in \Pcal$ comes from the Cartesian product of $\Qcal$
and $\Dcal$, namely
$\Pcal = \cbr{\rbr{q, d^{+}} ~|~ q \in \Qcal ~\text{and}~ d^{+}
  \in \Dcal}$ where $\Pcal \subseteq \Qcal \times \Dcal$. We may also be given, implicitly or explicitly, a set of
negative observations $\Ncal$, which also comes from the Cartesian
product of $\Qcal$ and $\Dcal$. Typically
$\abr{\Pcal} \ll |\Qcal \times \Dcal|$ and
$\Pcal \cap \Ncal = \emptyset$. We will use $\rbr{q, d^{+}}$ or
$\rbr{q_{i}, d_{j}^{+}}$ to denote elements of $\Pcal$, and
correspondingly represent elements of $\Ncal$ as $\rbr{q, d^{-}}$ or
$\rbr{q_{i}, d_{j}^{-}}$.

A dyadic embedding model, in turn, is a function that maps elements from
$\Qcal$ or $\Dcal$ into an $l$-dimensional Euclidean space endowed with
the usual Euclidean dot product $\inner{\cdot}{\cdot}$.
Broadly speaking, there are two types of dyadic embedding models: 
\begin{description}
\item[Factorized Models] where the embeddings $\qb$ and $\db$ of $q$
  and $d$ respectively are computed independently, via a function
  $f\rbr{\cdot}$. The training objective is chosen to ensure that
  $\inner{\qb}{\db^{+}} > \inner{\qb}{\db^{-}}$ for any pair of positive and negative documents $\db^{+}$ and $\db^{-}$. As can be seen, these
  models remain agnostic to the interrelation between the
  inputs. Examples of such models include the influential Deep Structured
  Semantic Model (DSSM) of \citet{huang2013dssm} as well as many others
\citep{mitra2017neural,shen2014latent,palangi2016deep,nigam2019semantic}.
  
\item[Interaction Models] where we compute joint embeddings of the form
  $g\rbr{q, d}$ where $g: \Qcal \times \Dcal \to \RR$
  is an embedding function. Clearly, such models take the relationship
  between $q$ and $d$ into account when determining vector
  representations \citep{guo2016deep, pang2016text,
    Hu:2014:CNN:2969033.2969055,
    hui2017pacrr,hui2017re,hui2018co,wan2016match,mitra2017learning}.
\end{description}
In this paper we focus exclusively on factorized models, simply because
they can be deployed at scale; we precompute the embeddings for all
elements of $\Dcal$, and given a query $q$ we simply need to compute
$f(q) := \qb$, and search for its nearest neighors in the set
$\cbr{\db_{1}, \ldots, \db_{m}}$ in the Euclidean space $\RR^{l}$. 

\begin{figure}
	\centering
	\includegraphics[height=5cm,width=5cm]{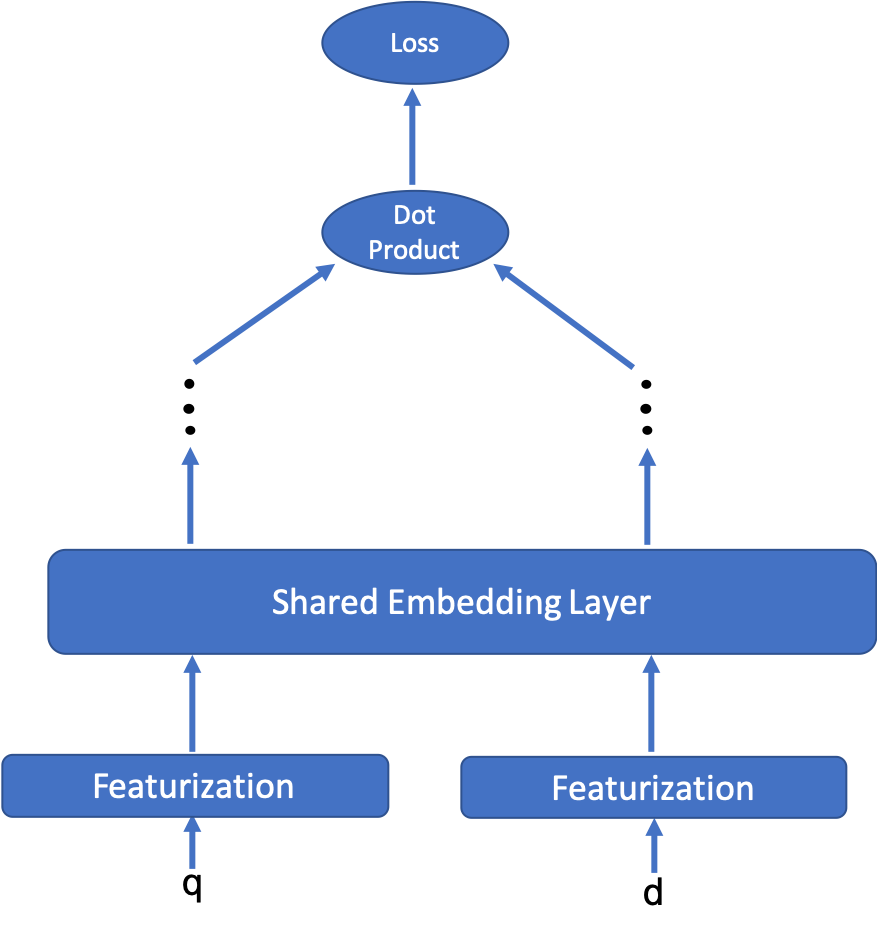}
	\caption{High-level factorized model architecture; one can also use separate embedding layers for $q$ and $d$ instead of sharing parameters.}
  \label{fig:modelarch}
\end{figure}

In Figure~\ref{fig:modelarch}, we present a high-level schematic of a
factorized model architecture, which takes the form of a Siamese network
that computes embeddings for two inputs using a deep neural network,
before calculating a dot product and loss. Although other variants are
possible (e.g., by replacing the dot product with a different similarity
function), we will work with this prototypical model in this
paper. Moreover, we will assume that the loss is computed in a
pointwise manner. Again, one can work with a variety of loss functions,
but, for simplicity, we will only focus on the squared hinge loss:
\begin{equation}
\label{eqn:squared_hinge} 
L(\hat{y}, y) := y \cdot \min(0, \hat{y}-t_1)^2 + (1-y) \cdot \max(0, \hat{y} - t_2)^2 
\end{equation}
where $y=1$ if a pair is positive and 0 otherwise and $t_1$ and $t_2$ denote the thresholds for positive and negative examples, respectively.  

\section{Scaling up Training and Inference}
\label{sec:SpeedingupTraining}

\subsection{Preprocessing and Graph Clustering}
\label{sec:PreprGraphClust}

\begin{figure}[h]
	\centering
	\includegraphics[width=\linewidth]{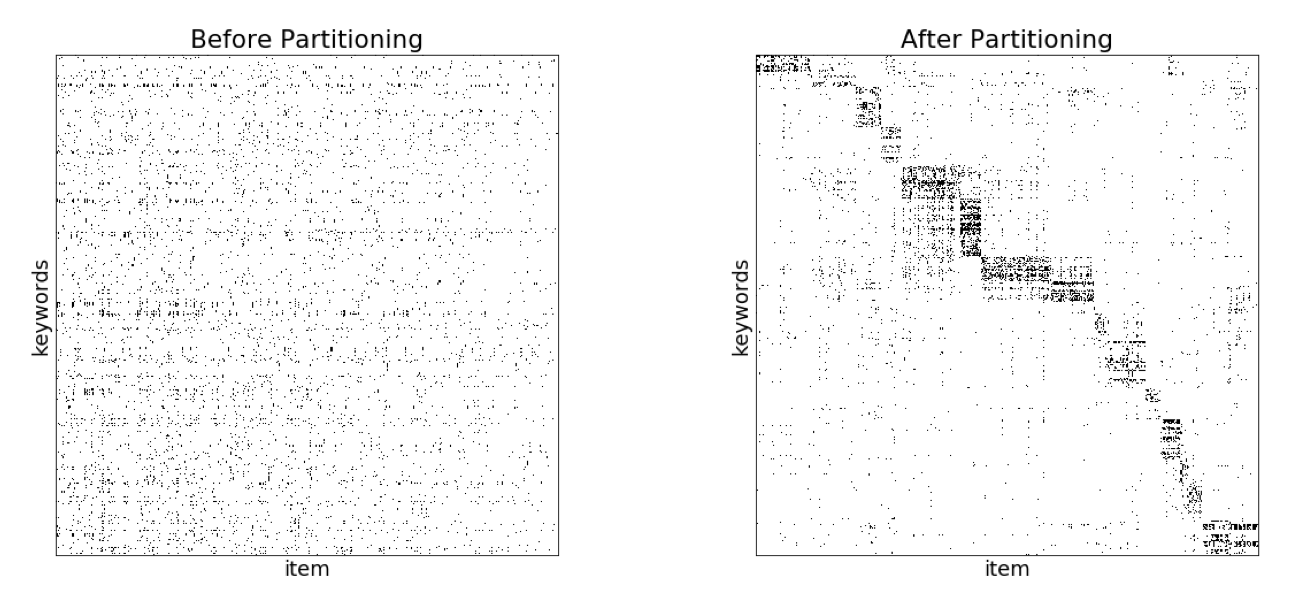}
	\caption{Co-occurence matrix of queries and items in a product search dataset. Left: Co-occurence before partitioning where dark points
	indicate a purchase. Right: Co-occurence matrix after reordering queries and items by the partitioning}
  \label{fig:metis}
\end{figure}

\begin{figure}[ht]
	\centering
	\includegraphics[width=\linewidth,height=5cm]{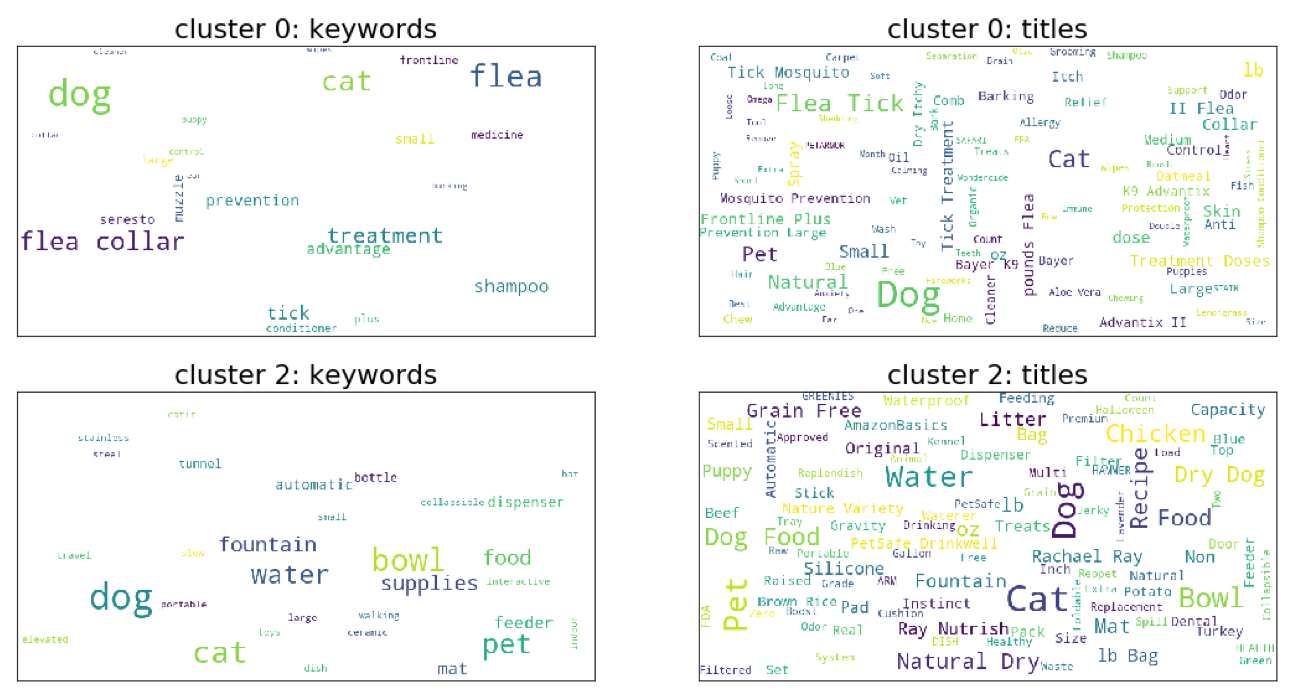}
	\caption{Word cloud of the frequent terms in two different clusters in the pets category of an e-commerce dataset. Cluster 0 corresponds to
	dog flea treatments while Cluster 2 centers on dog and cat food}
  \label{fig:wordcloud}
\end{figure}

We use the METIS library \cite{karypis1998fast} to cluster the bipartite graph derived from the dyadic data. As shown in Figure \ref{fig:metis}, our real-world product search dataset is highly structured, with the partitioning identifying a clear block-diagonal structure in the co-occurence matrix of queries and items. Moreover, Figure \ref{fig:wordcloud} depicts that these clusters also contain a semantic coherence that is distinct from other partitions by plotting the frequent terms in the queries and product titles within two sample clusters. In the product search dataset used in our experiments, the graph edges represent purchased products in response to a query, weighted by the number of purchases. METIS also enforces a \emph{balance} between clusters, stipulating that each cluster has roughly the same number of nodes (either queries or products). The balance property is especially important for our inference algorithm since we would like to build the indexes and perform the nearest neighbor search within a given partition quickly and therefore avoid degenerate clusters containing a large fraction of items. Due to the importance of balance, we favor algorithms like METIS over other types of clustering approaches such as $k$-means clustering over the embeddings directly. However, we note that METIS, as applied to our bipartite graph, enforces a balance only between the union of queries and documents; we may still observe some variance in the number of documents per partition as shown in Figure \ref{fig:partition_size_hist}.

\subsection{Training}
\label{sec:Training}

In this section, we discuss our proposed negative sampling technique in further detail. Let $\Gcal = (\Qcal \cup \Dcal, \Pcal)$ be a bipartite graph derived from our training set where an edge $(q, d) \in \Pcal$ if and only if $q \in \Qcal, d \in \Dcal$ and $(q, d)$ has a positive association. Let $\Ccal$ denote a partition of the vertices of $G$ into $r$ clusters $c_1, \dots, c_r$ such that each vertex in $G$ belongs to one and only one partition\footnote{We chose to uniquely assign each node to a single partition to reduce memory, but, in principle, one could replicate entities across clusters. We leave this exploration for future work.}.

Given access to such a partitioning, we propose Algorithm \ref{alg:alg1} to sample negative examples for a minibatch during training. 

 \begin{algorithm}[h]
    \caption{Hard Negative Mining via Graph Partitioning \newline
    \textbf{Input:} \text{Partitions $\{c_1, \dots, c_r\}$, window size $w$, sample size $t$}, and  \newline
    \text{queries $\{q_1, \dots, q_n\}$}
    }
    \begin{algorithmic}[1]
    \STATE $S \gets \{\}$
    \FOR{$q_i$ in $\{q_1, \dots, q_n\}$}
    \STATE Look up cluster $c_i$ containing $q_i$. 
    \STATE Get the top $w$ partitions $W$ by cluster affinity with $c_i$. 
    \STATE Select a high-affinity cluster $c_j$ uniformly at random from $W$, excluding $c_i$. 
    \STATE Sample $\lceil t/n \rceil$ documents $d_{i1}^-, d_{i2}^-, \dots, d_{is}^-$ uniformly at random from $c_j$
    \STATE $S \gets S \cup \{(q_i, d^-_{i1}), \dots, (q_i, d^-_{is})\}$
    \ENDFOR
    \RETURN $S$
    \end{algorithmic}
    \label{alg:alg1}
    \end{algorithm}
    
In Algorithm \ref{alg:alg1} we can utilize various definitions of cluster affinity. In our work, we rely on the number of edges that cross between two clusters as a measure of their affinity. Intuitively, these edge cuts measure affinity as we expect to see more overlap between clusters pertaining to men's and women's shoes than, say, men's shoes and dog food. In our experiments, we found that uniformly sampling from a fixed number of top clusters as opposed to selecting clusters with probability proportional to their affinity provided better model performance. We hypothesize that this phenomenon is due to the importance of \emph{diversity} in our samples. In particular, uniform sampling allows us to include negative samples from a variety of clusters whereas a probability distribution based on cluster affinity tends to favor only the top clusters. We note that one might also be able to extend this algorithm into a natural curriculum learning scheme where we progressively tighten the window parameter $w$ over the course of training. We defer this investigation for future work. 

\subsection{Inference}
\label{sec:Inference}

Let $\Gcal = (\Qcal \cup \Dcal, \Pcal)$ be a bipartite graph constructed from a set of positively associated query-document pairs as defined in the previous section and let $\Ccal$ denote a partition of $\Gcal$ into clusters $p_1, \dots, p_c$.
We propose to use this partitioned graph for a more scalable approximate $k$-nearest neighbors algorithm as follows: we will train a classifier that, given a query embedding $\qb$, predicts the clusters with the highest affinity to $\qb$. We then perform a nearest neighbor search inside these clusters to return our final result using a backend KNN algorithm $\mathcal{A}$ of our choice. As an additional optimization, we introduce a cumulative probability cutoff where we will stop probing for additional clusters if the cumulative probability of the clusters we have visited thus far, as predicted by our classifier model, exceeds a provided threshold of $t$. 

\begin{algorithm}[h]
    \caption{Partitioned Nearest Neighbor Search (PNNS) \newline
    \textbf{Input:} \text{Partitions $\{c_1, \dots, c_r\}$, query embedding $\qb$, classifier $h$,} \newline
    \text{number of probes $d$, number of neighbors $k$, probability cutoff $t$,} \newline 
    \text{and a backend KNN algorithm $\mathcal{A}$}
    }
    \begin{algorithmic}[1]
    \STATE Compute $s_i = h(q, c_i)$  for $i \in \{1, ..., d\}$
    \STATE Identify the top $w$ clusters $c^{\prime}_1, ... c^{\prime}_w$ where \newline $s^{\prime}_i \ge s_{i+1} \forall i \in \{1,...d\}$, and $\sum_{i=1}^{w} s’_i \ge t$
    \RETURN $\mathcal{A}(k, [c^{\prime}_1, \dots, c^{\prime}_w])$, the $k$ nearest neighbors computed by the backend algorithm across
    the top $w$ clusters. 
    \end{algorithmic}
    \end{algorithm}

In our experiments, we perform the nearest neighbor search over our candidate clusters serially. One could also perform the search over in each cluster in parallel and reduce the search latency further. We defer this optimization for future work. 

\subsection{Cluster Prediction Models}
\label{sec:ClustPredModels}

Our cluster prediction model takes a query embedding vector $\qb$ as input and outputs a probability distribution over all of partitions, representing the likelihood of a given cluster containing relevant documents to the query. In our experiments, we use a two-layer feed forward neural network followed by a softmax layer with 256 hidden nodes in each hidden layer and a crossentropy loss. We train the model on a set of query vectors computed by an embedding model and supervise over the labeled cluster containing the query. 

\begin{figure}[h]
	\centering
	\includegraphics[width=\linewidth]{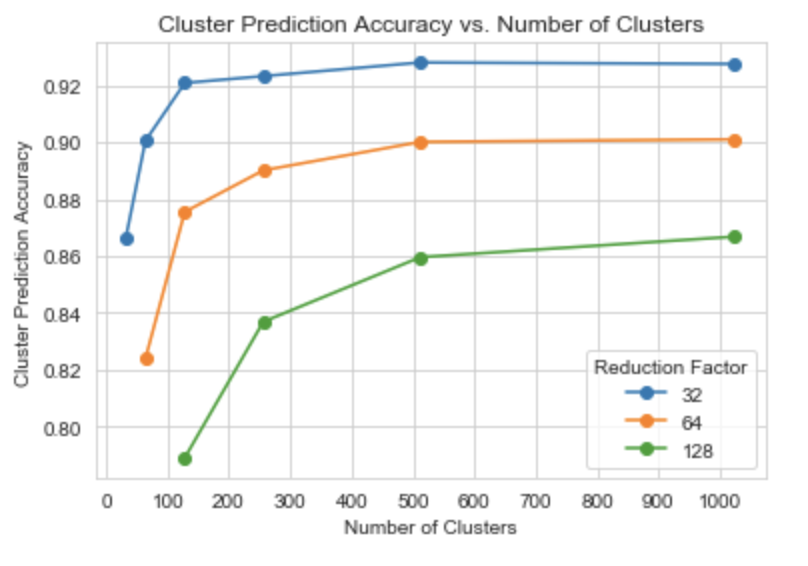}
	\caption{Evaluation of the tradeoff between number of partitions and number of probes in PNNS. The ``reduction" factor represents the ratio between the number of clusters to the number of probes. We observe that the classifier accuracy increases as we examine more clusters (since, for a fixed reduction factor, we add more probes). However, the accuracy eventually plateaus, which suggests that the underlying graph partitioning introduces some degree of noise in failing to group all relevant products together. }
  \label{fig:cluster_pred}
\end{figure}

We note that our partitioned nearest neighbors algorithm introduces two distinct sources of error: 1) the cluster prediction model could make an incorrect prediction and lead us to search in the wrong partitions and 2) the graph partitioning itself might fail to group certain relevant documents together. In Figure \ref{fig:cluster_pred} we plot the accuracy of our cluster prediction models in selecting the correct cluster for our test set of queries across different numbers of clusters and different numbers of probes. We define the ``reduction factor" as the ratio of the number of clusters to the number of probes to examine the tradeoff between searching in fewer clusters and the prediction accuracy. From these plots, we see that our prediction model suffers in performance with a larger reduction factor, which introduces a tradeoff between search latency which naturally decreases when we examine a smaller fraction of clusters. 

\section{Related Work}
\label{sec:RelatedWork}

\subsection{$k$-Nearest Neighbors}
\label{sec:kNearestNeighbors}

Exact and approximate KNN remains a fundamental algorithmic task that
has seen an increased interest with the advent of neural embedding
models. In the 1970s, \citeauthor{bentley1975multidimensional}
introduced the KD-tree, a data structure for dividing the Euclidean
space to enable efficient exact searching
\cite{bentley1975multidimensional}. However, KD-trees scale poorly with
respect to the dimension, and are therefore not suitable for most modern applications.

As opposed to KD-trees, which divide the Euclidean space by using a
data-structure, Locality-sensitive hashing (LSH) is an alternate
technique for \emph{approximate} KNN, which uses randomization to
quantize the space
\cite{indyk1998approximate,gionis1999similarity,andoni2006near}.
More recently, additional powerful approximate KNN algorithms have
emerged including the Hierarchical Navigable Small World (HNSW)
\cite{malkov2018efficient}, product quantization
\cite{jegou2010product}, cell probe methods such as the inverted
file index \cite{sivic2003video}, Navigating Spread-out Graph (NSG)
\cite{FuNSG17},  and Neighborhood Graph and Tree (NGT) \cite{iwasaki2015ngt}
along with libraries implementing these approaches such
as NMSLIB \cite{DBLP:conf/sisap/BoytsovN13}, FAISS \cite{JDH17}, and
Annoy \cite{bernhardsson2017annoy}.

The above techniques work in a data-independent manner. In contrast, there
is an exciting line of work recently that focuses on learned indices
\cite{KraBeuChiDeaetal18}. \citet{dong2019learning}
proposed a data-dependent algorithm for approximate KNN, which they call
Neural-LSH (also see \citep{sablayrolles2018spreading}). Applied to our
context, the algorithm works as follows: given the document embeddings
$\cbr{\db_{1}, \ldots, \db_{m}}$ construct a KNN graph, that is, link
$\db_{i}$ with $\db_{j}$ if $\db_{j}$ is a $k$-nearest neighbor of
$\db_{i}$. Given the KNN graph, find a balanced partition of this graph
by minimizing edge-cut. Finally, train a neural network classifier to
map document embeddings $\db$ to the corresponding graph partition. At
inference time, use the classifier to map the query embedding $\qb$ to a
partition and perform exact nearest neighbor search within that
partition. While this approach also leverages learning and graph partitioning 
to improve upon
classical techniques, the key difference between this algorithm and our
proposed solution lies in the former technique having to build a KNN graph, which requires performing a KNN
search for each point in the space. This operation proves to be prohibitively expensive on large datasets with hundreds of
millions or billions of points. In contrast, our method relies upon a 
graph that has already been constructed from our dyadic dataset, which eliminates
the need to build the network ourselves, saving hours, if not days, of compute time.

\subsection{Partitioning}
\label{sec:Partitioning}


We note in the passing that the problem of graph partitioning is NP-Hard
\cite{andreev2006balanced}. However, several approximation algorithms
such as METIS \cite{karypis1998fast}, KaHIP \cite{sandersschulz2013}, SCOTCH \cite{pellegrini1996scotch}, and
PuLP \cite{slota2014pulp} have been developed. We evaluated these methods on our product search dataset and 
settled on METIS for our experiments since it offered the best tradeoff between quality (recalling the relevant products
for a given query) and speed (partitioning our graph in roughly 6 hours). 

\subsection{Negative Sampling}
\label{sec:NegativeSampling}

A number of papers have explored principled approaches for identifying
informative training examples to decrease time to convergence for
training neural networks. A common theme in this body of work centers on
importance sampling, constructing a distribution over
training examples with greater weight given to samples more likely to
produce large parameter updates \cite{gao2015active, johnson2018training, guo2018vse}. 
These approaches, however, require
maintaining a distribution over all training examples, which becomes
infeasible at larger scales. In contrast, by clustering the data, we can
maintain a coarse-grained distribution over the clusters instead of each
training data point. A graph based approach for negative sampling, but
very different from ours, was proposed by \citet{ying2018graph} in the
context of the Pinsage algorithm. 


\section{Experiments}
\label{sec:Experiments}

In this section, we present experimental results demonstrating the scalability properties of our proposed partitioning scheme on a large-scale product search dataset. In particular, we focus on using graph partitioning for improving the training of embedding models through hard negative sampling and for efficiently deploying popular approximate KNN algorithms. We do not focus on presenting end-to-end retrieval results and instead focus on the improvements to training and deployment separately as the improvements to these sub-components can be applied independent of each other and can be extended to other dyadic data applications. In addition, we measure recall in comparison to the baseline of an exact KNN search, which is a relative measure and thus independent of any improvements to the underlying embedding model. 

\subsection{Data \& Algorithms}

We evaluate our proposed negative sampling approach using a product search dataset sampled from Amazon.com search logs. Our training set consists of tens of millions of unique search queries and products and hundreds of millions of training examples. We use the semantic product search model architecture proposed in \cite{nigam2019semantic} to learn query and product embeddings.

To evaluate our inference algorithm, we construct a dataset of product embeddings at the billion scale and use METIS to partition our data into 64 clusters. We benchmark the performance of KNN algorithms on a CPU machine against a set of 1000 query embeddings. In our experiments, we measure 1) the algorithm's ability to recall the 100 closest vectors for each query, 2) the average latency of a single query search, and 3) the time required to construct the approximate KNN index. We investigate scaling 3 popular algorithms with PNNS: HNSW (HNSWLIB implementation\footnote{https://github.com/nmslib/hnswlib}), NGT\footnote{https://github.com/yahoojapan/NGT}, and the Inverted File Index (IVF) method (Faiss implementation\footnote{https://github.com/facebookresearch/faiss}). For all algorithms, we use cosine similarity as our metric of choice to match with the similarity measure used to train these embeddings.

\subsection{Hardware}

We trained our embedding models on a single AWS p3.16xlarge machine with 8 NVIDIA Tesla V100 GPUs (16GB), Intel Xeon E5-2686v4 processors, and 488GB of RAM. 

We performed the METIS graph clustering as well as all KNN benchmarking experiments on an AWS x1e.32xlarge machine with 128 vCPUS, 4TB of memory, and quad socket Intel Xeon E7 8880 processors. 


\subsection{Negative Sampling Experiments}
\label{sec:NegativeSampling-2}

In this section, we present experimental results with our proposed negative sampling algorithm. As mentioned, we conduct all of our experiments with an embedding model tuned for product search. We construct a vocabulary consisting of 125,000 of the most frequent word unigrams, 25,000 word bigrams, and 50,000 character trigrams along with 500,000 additional tokens reserved for out-of-vocabulary terms, which we randomly hash into these bins. The inputs to our model are query keywords and product title text, which we tokenize into 32 and 128-length arrays from our vocabulary, respectively. We set our embedding dimension to 256, batch size to 8192, use Xavier weight initialization, and train using the Adam optimizer \citep{kingma2014adam} with $\alpha=0.001, \beta_1 = 0.9, \beta_2=0.999, \epsilon=10^{-8}$ and the aforementioned squared hinge loss function (Equation \ref{eqn:squared_hinge}) with thresholds $t_1=0.9$ and $t_2=0.2$. 

Since we are focused on ad hoc retrieval, we evaluate model performance on a hold-out validation test according to ``Matching" Mean Average Precision (MAP) and ``Matching" Recall as defined in \cite{nigam2019semantic} where we first sample a set of 20,000 queries and evaluate the model’s ability
to retrieve purchased products from a sub-corpus of 1 million products for those queries. 

Tables 1 and 2 show the result of our parameter sweep on our evaluation set for our proposed negative sampling algorithm where each row corresponds to the number of graph clusters used while each column represents the number of nearby clusters probed for samples. We observe that we hit diminishing returns with too many clusters where we might split relevant items into different partitions and, consequently, sample related pairs as negatives. Similarly, we notice that increasing the number of probes improves model performance, which we hypothesize is due to sampling a greater \emph{diversity} of negatives. However, increasing the number of probes comes at the cost of longer training times as we spend more computation within each training step sampling negatives.

\begin{table}[ht]
  \footnotesize
  \centering
  \begin{tabular}{c|c||*{8}{c|}}
    \multicolumn{2}{c}{} & \multicolumn{8}{c}{} \tabularnewline
    \cline{2-10}
    \multirow{8}*{\rotatebox{90}{}} &
&    \bfseries 8 & \bfseries 16 & \bfseries 32 & \bfseries 64 &\bfseries 128 & \bfseries 256 &\bfseries 512 &\bfseries 1024 \tabularnewline[1 ex] 
\cline{2-10}
&    \bfseries 2048 & 0.317 & 0.318 & 0.319 & 0.314 & 0.312 & 0.304 & 0.295  & 0.285 \tabularnewline [1ex] 
    \cline{2-10}
&    \bfseries 4096 & 0.323 & 0.326 & 0.327 & 0.321 & 0.320 & 0.312 & 0.306& 0.293  \tabularnewline [1ex] 
    \cline{2-10}
&    \bfseries 8192 & 0.328 & 0.330 & 0.339 & 0.331 & 0.332 & 0.321 & 0.312 & 0.302 \tabularnewline [1 ex]
    \cline{2-10}
&    \bfseries 16384 & 0.329 & 0.333 & 0.338 & 0.336 & 0.338 & 0.332 & 0.319 & 0.309  \tabularnewline [1 ex]
    \cline{2-10}
&    \bfseries 32768 & 0.323 & 0.332 & 0.338 & 0.337 & 0.334 & 0.339 & 0.328  & 0.310 \tabularnewline [1 ex]
    \cline{2-10}
&    \bfseries 65536 & 0.306 & 0.322 & 0.332 & 0.334  & 0.335 & \bf{0.341} & 0.331 & 0.307 \tabularnewline [1 ex]
    \cline{2-10}
&    \bfseries 131072 & 0.286 & 0.302 & 0.327 & 0.331 & 0.338 & 0.337 & 0.329 & 0.298 \tabularnewline [1 ex]
    \cline{2-10}
  \end{tabular}
  \caption{Match MAP across various number of clusters (rows) and number of sampling probes (columns)}
  \label{table:ns_map_sweep}
\end{table}

\begin{table}[ht]
  \footnotesize
  \centering
  \begin{tabular}{c|c||*{8}{c|}}
    \multicolumn{2}{c}{} & \multicolumn{8}{c}{} \tabularnewline
    \cline{2-10}
    \multirow{8}*{\rotatebox{90}{\centering{}}} &
&    \bfseries 8 & \bfseries 16 & \bfseries 32 & \bfseries 64 &\bfseries 128 & \bfseries 256 &\bfseries 512 &\bfseries 1024  \tabularnewline[1 ex] 
\cline{2-10}
    \cline{2-10}
&    \bfseries 2048 & 0.761 & 0.775 & 0.780 & 0.778 & 0.784 & 0.780 & 0.772 & 0.761 \tabularnewline [1ex] 
    \cline{2-10}
&    \bfseries 4096 & 0.754 & 0.767 & 0.777 & 0.781 & 0.784 & 0.783 & 0.782 & 0.770 \tabularnewline [1ex] 
    \cline{2-10}
&    \bfseries 8192 & 0.739 & 0.757 & 0.775 & 0.782 & 0.790 & 0.787 & 0.789 & 0.778 \tabularnewline [1 ex]
    \cline{2-10}
&    \bfseries 16384 & 0.724 & 0.747 & 0.762 & 0.773 & 0.786 & 0.788 & \bf{0.790} & 0.782 \tabularnewline [1 ex]
    \cline{2-10}
&    \bfseries 32768 & 0.703 & 0.723 & 0.746 & 0.757 & 0.772 & 0.787 & 0.786 & 0.778 \tabularnewline [1 ex]
    \cline{2-10}
&    \bfseries 65536 & 0.672 & 0.697 & 0.715 & 0.743 & 0.763 & 0.775 & 0.782 & 0.771 \bfseries \tabularnewline [1 ex]
    \cline{2-10}
    &    \bfseries 131072 & 0.635 & 0.661 & 0.696 & 0.720 & 0.743 & 0.760 & 0.768 & 0.754 \bfseries \tabularnewline [1 ex]
    \cline{2-10}
  \end{tabular}
  \caption{Match Recall across various number of clusters (rows) and number of sampling probes (columns)}
   \label{table:ns_recall_sweep}
\end{table}

\begin{figure}[h]
	\centering
	\includegraphics[width=\linewidth]{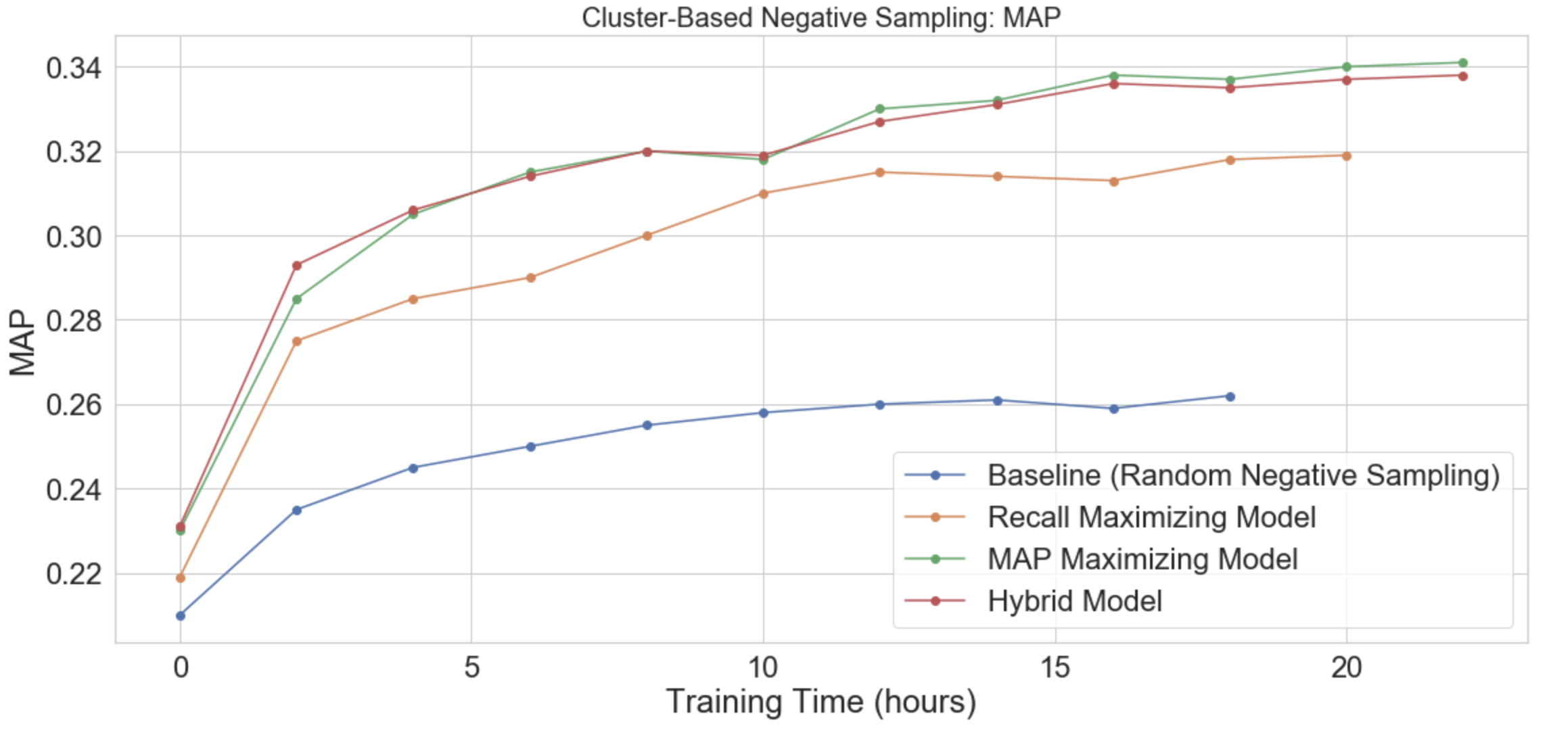}
	\caption{Relative time plot of Matching MAP }
  \label{fig:ns_map_plot}
\end{figure}

\begin{figure}[h]
	\centering
	\includegraphics[width=\linewidth]{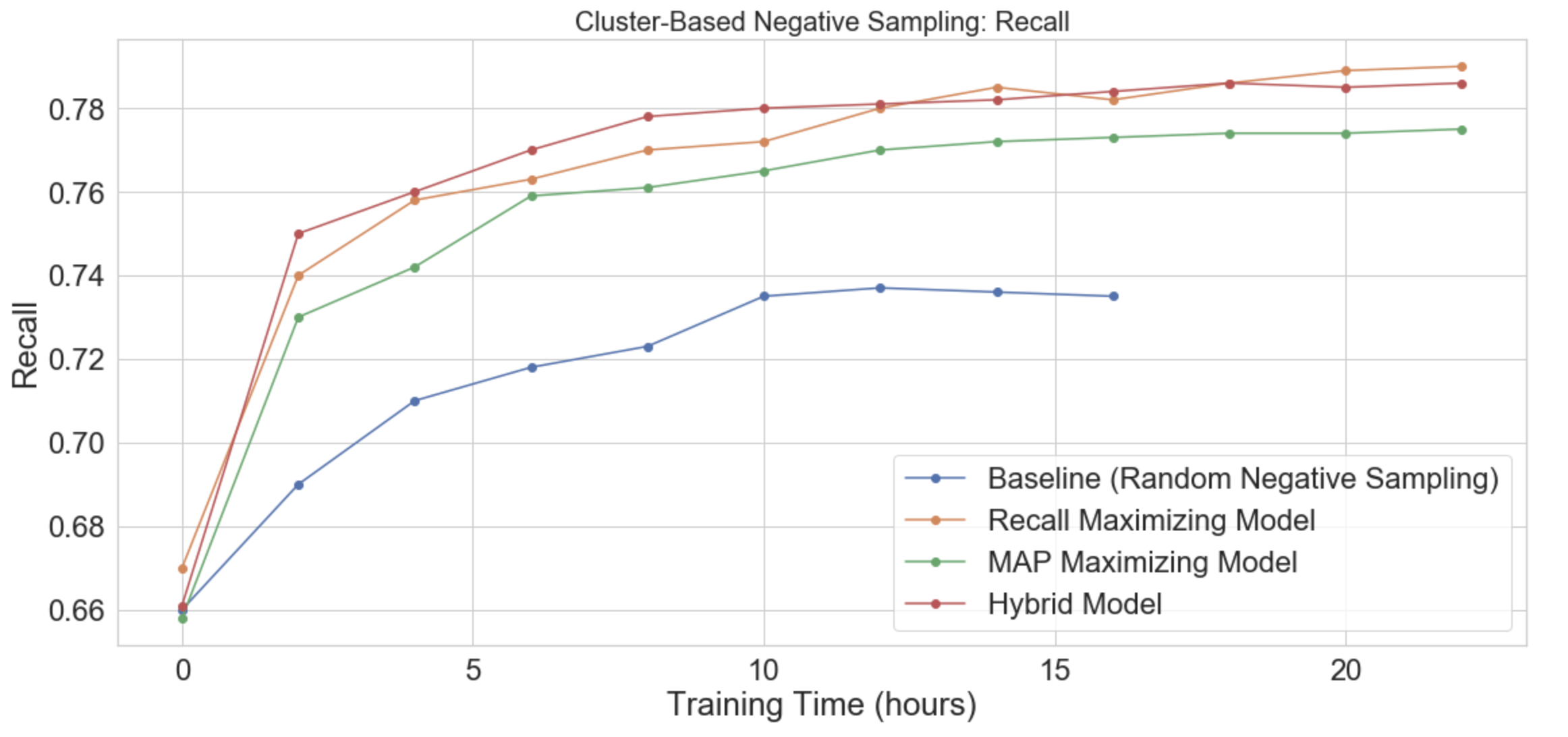}
	\caption{Relative time plot of Matching Recall }
  \label{fig:ns_recall_plot}
\end{figure}

Secondly, we can compare our best performing graph-based negative sampling models to our baseline with random negative sampling. Based on Tables \ref{table:ns_map_sweep} and \ref{table:ns_recall_sweep}, we select the model with the best performing MAP (65,536 clusters and 256 probes), the model with the best Recall (16,384 clusters and 512 probes) and a hybrid model that achieves strong performance on both metrics (16384 clusters, 128 probes). In Figures \ref{fig:ns_map_plot} and \ref{fig:ns_recall_plot}, we compare these models to a baseline which sampled negatives uniformly at random while keeping all other parameters fixed. In these plots, we compare the relative training times for each model by measuring metrics across hours of training time.  Since the baseline involved no computation between each minibatch aside from uniform random sampling, each step of the baseline was approximately twice as fast as each step of the graph-based sampling models. However, these graph-based sampling models compensated for their added computation per step. They generalize better on the test set and achieve stronger performance on our validation metrics for every fixed unit of time past the start of training, allowing us to train a better model in less time than the baseline.

\subsection{KNN Experiments}
\label{sec:KNN}

We next turn our attention to benchmarking the performance of our Partitioned Nearest Neighbor Search (PNNS) algorithmic framework for scaling KNN search on a billion-scale collection of vectors. We conducted grid searches on a smaller collection size of 3 million vectors to identify performant hyperparameter settings that achieved over 95\% recall for each algorithm. Ultimately, we settled on the following parameter settings for our experiments:

\begin{itemize}

\item NGT: ESC=30, ESS=70 (ESC=10, ESS=20 with no partitioning)

\item HNSW: EFC=700, EF=700, M=110

\item IVF: NLIST=256, NUM PROBES=16

\end{itemize}

We note that we selected weaker hyper-parameter settings for NGT without partitioning because we observed that the algorithm would take an intractably long time (at least several months) to build the index otherwise. However, with PNNS partitioning, we were able to use more aggressive hyperparameters to achieve comparable latency and recall results to the other algorithms used in our experiment. 

\subsubsection{Index Build Time}

One challenge with deploying approximate KNN algorithms at the billion scale is the fact that these approaches almost always involve a time-consuming step of converting input vectors into an index structure for searching. Many production search systems elect to rebuild their indexes at a regular cadence, such as every 24 hours. Although approximate approaches often dramatically reduce the search latency relative to a brute force search at marginal losses of recall, the index build time can possibly take multiple days, making daily rebuilds of the index infeasible. Through the PNNS graph partitioning approach, we can reduce this build time by building the indexes for each partition in parallel across multiple machines. Such a multi-machine index build is not currently supported by the libraries we experimented with.

\begin{figure}[h]
	\centering
	\includegraphics[width=\linewidth]{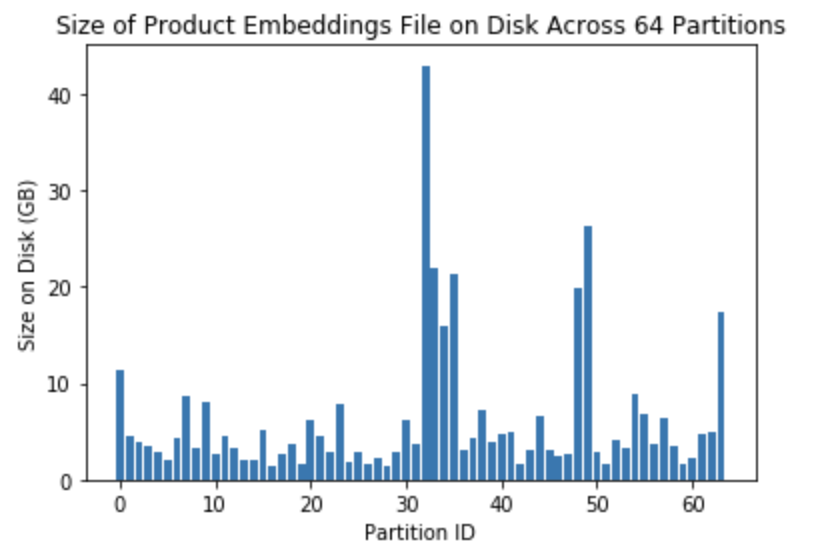}
	\caption{Size of the vector embeddings file across partitions. We observe that although METIS enforces a balance between the number of vertices across partitions in the bipartite graph, we may still have some imbalance when restricting our focus just to the documents.}
  \label{fig:partition_size_hist}
\end{figure}

Since we partition a bipartite graph of queries and products, the METIS algorithm enforces a balance between the sum of query and product nodes per cluster. Thus, we may still have a range in the number of products per cluster, as shown in Figure \ref{fig:partition_size_hist}. As a result, we find that the overall index build time does not scale purely linearly with the number of machines since some of the partitions take longer to build than others because they contain more documents. This problem of efficiently building the indexes for each partition across some number of machines is an instance of the classic algorithmic task of assigning jobs to machines. For simplicity, we employ the well-known greedy algorithm of first sorting the jobs by their respective amounts of work and then iteratively assigning the most intensive remaining job to the machine currently with the lightest load. This approach guarantees an assignment of jobs where the maximum load across all machines is at most a constant factor of $4/3$ more than the optimal solution, as first shown in the classical paper of \citet{graham69boundson}. 

In Table \ref{table:pnns_index_build}, we report the index build time both for the standard KNN algorithms with no partitioning and with PNNS across a various number of machines. To conserve computational resources, we simulate running the PNNS index build over multiple machines by only running the jobs assigned to the machine with the maximum load, which will determine the overall index build time. From these results, we see that PNNS can reduce the cost of the overall index build time and, in some cases, make daily index builds feasible when such a cadence would not be possible without partitioning. We note that the times reported in Table \ref{table:pnns_index_build} do not include the additional computation required for the graph partitioning. However, we can avoid re-running the partitioning step on a daily basis by assigning new documents to clusters via our classifier. Thus, in an amortized sense, the cost of graph partitioning becomes negligible compared to the cost of building the KNN indexes. 

We also note that the savings in the build time from our partitioning scheme do come at the cost of increased computational resources as we construct the KNN indexes in parallel. However, since none of the algorithms/libraries we benchmark currently enable multi-machine index building, they cannot scale in the same manner with more compute resources. Thus, our method provides an avenue for indexing billion-scale embedding data for popular KNN search approaches within the constraints of a product build cadence, such as daily updates. 

\begin{table}[ht]
\small
  \centering
  \begin{tabular}{c|c|*{5}{c|}}
    \cline{2-5}
    \multirow{8}*{\rotatebox{90}{\centering{\textbf{}}}} &
 & \bfseries NGT & \bfseries HNSW & \bfseries IVF \tabularnewline [1 ex] 
\cline{2-5}
&    \bfseries No Partitioning &  >1000 &  87.2 & 9.28   \tabularnewline [1ex] 
    \cline{2-5}
&    \bfseries PNNS (2 machines) & 77.03  & 33.37    &   2.12 \tabularnewline [1ex] 
    \cline{2-5}
&    \bfseries PNNS (4 machines) & 38.35   &   18.15 & 1.03   \tabularnewline [1ex] 
    \cline{2-5}
    &    \bfseries PNNS (8 machines) & 23.5   &   12.13 &    0.533  \tabularnewline [1ex] 
    \cline{2-5}
    &    \bfseries PNNS (16 machines) &  21.07  & 11.42    &   0.483 \tabularnewline [1ex] 
    \cline{2-5}
    \cline{2-5}
  \end{tabular}
  \caption{Index Build Time (hours)}
\label{table:pnns_index_build}
\end{table}

 \subsubsection{Recall and Latency}
 
In this section we focus on benchmarking the recall and latency of PNNS on our billion-scale collection size. As mentioned, we define recall as $\frac{|S_E \cap S_A|}{|S_E|}$ where $S_E$ is the set of results returned by an exact KNN search and $S_A$ is the items retrieved by the approximate algorithm. In our experiments, we focus on recall@100, namely the ability of the approximate algorithm to retrieve the top 100 results returned by an exact search. In Table \ref{table:pnns_recall}, we report the average recall@100 across the 1000 queries we evaluated against. For latency, we measure the average time for each algorithm to return results across our 1000 benchmark queries.

In all of our PNNS experiments, we evaluate the algorithm's recall and latency across varying number of probes. In addition, we fix the cumulative cluster probability hyperparameter to 0.99, meaning that we terminate our search early if the cumulative probability of the clusters we have searched within, as predicted by our cluster prediction classifier, exceeds this threshold. 

\begin{table}[ht]
\small
  \centering
  \begin{tabular}{c|c|*{5}{c|}}
    \cline{2-5}
    \multirow{10}*{\rotatebox{90}{\centering{\textbf{Number of Probes}}}} &
 & \bfseries NGT & \bfseries HNSW & \bfseries IVF \tabularnewline [1 ex] 
\cline{2-5}
&    \bfseries 1  & 0.737  &  0.744 &  0.735 \tabularnewline [1ex] 
    \cline{2-5}
&    \bfseries 2 &  0.838 &  0.846  &  0.833 \tabularnewline [1ex] 
    \cline{2-5}
&    \bfseries 4 & 0.898 & 0.907 &  0.892 \tabularnewline [1ex] 
    \cline{2-5}
&    \bfseries 8 & 0.934 &  0.943 &  0.928 \tabularnewline [1ex] 
    \cline{2-5}
    &    \bfseries 16 & 0.957 & 0.967 & 0.950 \tabularnewline [1ex] 
    \cline{2-5}
    &    \bfseries \text{No Partitioning} & 0.756 & 0.980 & 0.983 \tabularnewline [1ex] 
    \cline{2-5}
  \end{tabular}
  \caption{PNNS Recall@100}
  \label{table:pnns_recall}
\end{table} 

\begin{table}[ht]
\small
  \centering
  \begin{tabular}{c|c|*{5}{c|}}
    \cline{2-5}
    \multirow{10}*{\rotatebox{90}{\centering{\textbf{Number of Probes}}}} &
 & \bfseries NGT & \bfseries HNSW & \bfseries IVF \tabularnewline [1 ex] 
\cline{2-5}
&    \bfseries 1  &  38.03 & 71.61   &  199.82  \tabularnewline [1ex] 
    \cline{2-5}
&    \bfseries 2 &  53.90 & 113.09    & 277.93   \tabularnewline [1ex] 
    \cline{2-5}
&    \bfseries 4 &  89.85 & 183.89   &  526.60  \tabularnewline [1ex] 
    \cline{2-5}
&    \bfseries 8 & 151.09  & 300.39 &   812.89  \tabularnewline [1ex] 
    \cline{2-5}
    &    \bfseries 16 & 289.12  & 453.69  &  1265.90   \tabularnewline [1ex] 
    \cline{2-5}
    &    \bfseries \text{No Partitioning} & 90.0 & 36.54  &  30313.07   \tabularnewline [1ex] 
    \cline{2-5}
  \end{tabular}
  \caption{PNNS Latency (ms)}
  \label{table:pnns_latency}
\end{table} 

From Tables \ref{table:pnns_recall} and \ref{table:pnns_latency}, we see that PNNS, for a larger fraction of probes, can achieve slightly reduced recall numbers compared to the standard HNSW algorithm without partitioning, but at the cost of increased latency. In practice, this tradeoff might still be favorable given the potential for a considerably decreased index build time as shown in Table \ref{table:pnns_index_build}. In addition, we find that PNNS can enable us to use more performant hyperparameter settings for NGT since the index build time becomes tractable. As a result, PNNS can provide a path for deploying NGT at the billion scale in practice and provides feasible latency and recall results across a variety of probes. The IVF algorithm has a significantly smaller index build time than the other methods we benchmarked, but comes at the cost of a much larger latency (though still orders-of-magnitude faster than an exact search). For IVF, we found that PNNS produced relatively marginal savings in index build time, but was able to reduce the search latency by an order of magnitude with a small reduction in recall. 

In summary, we found that PNNS can serve as a general algorithmic framework to run the popular HNSW and NGT approximate algorithms at the billion scale in practical settings by making daily index builds feasible. In the case of IVF, PNNS also reduced the search latency considerably. When assessing the tradeoffs between PNNS and the baselines techniques, we note that building the search index within 24 hours to facilitate daily updates may be an essential requirement for real production systems, in which case the ability of PNNS to parallelize the index build process may be essential at this scale despite the loss of some recall and a potential increase in latency. Furthermore, we note that each of the billion-scale KNN search indexes with no partitioning that we experimented with yielded a memory footprint of over 1 terabyte. While we were able to accommodate these large indexes in our experiments thanks to our use of an AWS x1 instance, we note that many practitioners might be looking for ways to deploy cheaper instances in production settings. The efficient partitioning strategy behind PNNS also provides a path for enabling distributed nearest neighbor search where we can store the indexes for each partition, which will have smaller memory footprints than the full index, over multiple machines and thereby sidestep this memory bottleneck. In this sense, PNNS, while possibly producing some regression in latency or recall when compared to its unpartitioned counterpart algorithm, may be an essential step in deploying nearest neighbor search at the billion scale within the practical constraints of real systems. 

\subsubsection{Deployment}

We successfully deployed PNNS-based search for several weeks in an online A/B test on a large e-commerce website to augment the traditional inverted index-based keyword matches with products retrieved by our neural embedding model. We validated that PNNS was able to meet the constraints of the search system while improving several business metrics. These results also validate that learned index structures can be integrated into established production systems and meet strict engineering requirements where data-independent algorithms fall short.

\section{Conclusion and Future Work}
\label{sec:ConclusionFutureWork}

In this paper, we address the problem of performing training and inference for dyadic embedding models at the billion scale, focusing on the practical application of semantic product search. To our knowledge, our work is the first to present solutions for deploying embedding-based retrieval at this scale under the constraints of realistic industrial systems. We demonstrated that the same underlying principle of leveraging the structure of real-world data can tackle both of these problems. By modeling dyadic data as a bipartite graph and utilizing balanced graph partitioning algorithms, we showed both improved model performance and reduced convergence time during training through efficient hard negative sampling. In addition, we presented a technique, based on graph clustering and a learned classifier, for scaling popular KNN algorithms in terms of either search time (in the case of IVF) or in sharply reducing the index build time (in the case of NGT and HNSW) with minimal impact on recall. Unlike similar graph partitioning approaches for KNN search in the literature, our technique leverages a graph already constructed from the underlying dyadic data and thereby eliminates the computationally prohibitive step of constructing a KNN graph at the billion scale. For future work on the negative sampling side, we can investigate curriculum learning strategies for our graph-based negative sampling approach where we tighten the window of adjacent cluster to sample from over the course of training.  On the inference side, as mentioned in \cite{dong2019learning}, we can consider investigating methods for learning the graph partitioning and the cluster prediction model in an end-to-end fashion where one optimization problem informs the other. Finally, we note that our proposed techniques are general in nature and can be applied to numerous problem domains that fit the dyadic data paradigm, and we hope to extend our ideas to other applications beyond product search. 



\bibliographystyle{ACM-Reference-Format}
\bibliography{graph_partitioning_paper}

%

\end{document}